\documentclass[12pt, notitlepage]{article}
\pdfoutput=1

\usepackage{float}
\usepackage{subfig}
\usepackage{graphicx}
\usepackage[T1]{fontenc}
\usepackage[utf8]{inputenc}
\usepackage{authblk}
\usepackage{amsmath,bm}
\usepackage[left=3cm,right=3cm,top=2cm,bottom=2cm]{geometry}
\renewcommand\footnotemark{}

\usepackage{verbatim}

\newcommand{\rf}[1]{(\ref{#1})}
\newcommand{\beq}{\begin{equation}}
\newcommand{\beql}[1]{\beq\label{#1}}
\newcommand{\eeq}{\end{equation}}
\newcommand{\bea}{\begin{eqnarray}}
\newcommand{\eea}{\end{eqnarray}}

\newcommand{\dd}{\mathrm{d}}

\newcommand{\eps}{\epsilon}

\newcommand{\cD}{{\cal D}}

\newcommand{\cM}{{\cal M}}

\newcommand{\cT}{{\cal T}}

\newcommand{\cZ}{{\cal Z}}

\usepackage{color}

\begin{document}

\title{Impact of topology in causal dynamical triangulations quantum gravity}         

\author[a,b]{J.~Ambj\o rn}
\author[c]{Z.~Drogosz}
\author[c]{J.~Gizbert-Studnicki}
\author[a,c]{A.~G\"orlich}
\author[c]{J.~Jurkiewicz}
\author[d]{D.~Nemeth}

\affil[a]{\small{The Niels Bohr Institute, Copenhagen University, \authorcr Blegdamsvej 17, DK-2100 Copenhagen Ø, Denmark. \authorcr E-mail: ambjorn@nbi.dk \vspace{+2ex}}} 

\affil[b]{\small{IMAPP, Radboud University, \authorcr Nijmegen, PO Box 9010, The Netherlands.\vspace{+2ex}}}

\affil[c]{\small{Institute of Physics, Jagiellonian University,  ul.~prof. Stanislawa Lojasiewicza 11,  PL 30-348 Krakow, Poland. \authorcr Email: zbigniew.drogosz@uj.edu.pl,  jakub.gizbert-studnicki@uj.edu.pl, andrzej.goerlich@uj.edu.pl, jerzy.jurkiewicz@uj.edu.pl\vspace{+2ex}}}

\affil[d]{\small{ Institute of Physics, 
E\"otv\"os Lor\'and University, P\'azm\'any P\'eter s\'et\'any 1/A, 1117 Budapest, Hungary
\authorcr Email: nemeth.daniel.1992@gmail.com}}

\date{\small({Dated: \today})}          
\maketitle


\begin{abstract}
 
We investigate the impact of spatial topology  in 3+1 dimensional causal dynamical triangulations (CDT) by performing numerical simulations with toroidal spatial topology  instead of the previously used spherical topology. {In the case of spherical spatial
topology we observed in the so-called phase C an average spatial volume distribution $n(t)$ 
which after a suitable time redefinition could be identified as the spatial volume distribution of the four-sphere. Imposing toroidal spatial topology we find that the average spatial 
volume distribution $n(t)$ is constant. By measuring the covariance matrix of spatial volume fluctuations we determine the form of the effective action. The difference compared
to the spherical case is that the effective potential has changed such that it allows
a constant average $n(t)$. This is what we observe and this is what one would
expect from a minisuperspace GR action where only the scale factor is kept as 
dynamical variable. Although no background geometry is put in by hand, the full 
quantum theory of CDT is also with toroidal spatial toplogy able to identify a classical background geometry around which there are well defined quantum fluctuations.
}

\vspace{1cm}

\end{abstract}


\section{Introduction}

An impressive progress of computing power during last twenty years  made numerical methods an important tools of contemporary  theoretical physics.  
One example is lattice QCD
which provides solutions for low-energy problems of strong interactions not tractable by means of analytic or perturbative methods. The progress has also been used in the quest for a theory of quantum gravity, where nonperturbative approaches play an important role. It is well known that  gravity cannot be formulated as a perturbative quantum field theory as it is  non-renormalizable \cite{tHooft:1974bx,Goroff:1985th}.
However, following Weinberg's  asymptotic safety conjecture \cite{Weinberg79}, there is hope that it can  be formulated as a predictive theory in a nonperturbative way. 
{The use of functional renormalization group equations have provided some
evidence in favor of the asymptotic safety conjecture in the case of gravity \cite{Kawai:1989yh,Kawai:1992np,Reuter:1996cp,Codello:2008vh,Reuter:2007rv,Niedermaier:2006wt,Litim:2003vp}, 
but since it involves approximations which are difficult to control, it is important 
to test the asymptotic safety scenario using other methods, and here the lattice 
models of gravity may play an important role. The starting point is a lattice regularization
of the gravity path integral}
\beql{Zcont}
\cZ= \int \cD_{\cM}{[g]} e^{iS_{HE[g]}}\ ,
\eeq
where one integrates over geometries, i.e. the equivalence classes of
spacetime metrics $g$ with respect to the diffeomorphism group on manifold $\cM$, and $S_{HE}$ is a classical gravitational action.
To give meaning to the formal expression \rf{Zcont}, one can introduce a lattice discretization. The finite lattice spacing $a$ provides a high
energy cut-off, and by taking $a\to 0$ one can in principle  approach  the continuum limit.
{Using what is known as dynamical triangulations (DT)
one starts out in the Euclidean sector, i.e. the path integral  \rf{Zcont}
is changed such that  one integrates over geometries with Euclidean signature 
and uses the corresponding Euclidean action ($iS_{HE}[g_L] \to 
-S_{HE} [g_E]$, where $g_L$ denotes a geometry with Lorentzian signature and 
$g_E$ denotes a geometry with Euclidean signature). The regularized (Euclidean)
path integral is then} 
\beql{Zdiscr}
\cZ = \sum_{\cT}e^{-S_R[\cT]}\ ,
\eeq
where the sum is over (abstract) triangulations $\cT$. {Thus  each abstract triangulation is viewed as a piecewise linear geometry where each link is assigned the length  $a$}, 
 and $S_R$ is the discretized Hilbert-Einstein action obtained  following Regge's method for describing piecewise linear geometries \cite{Regge:1961px}. Such a partition function  can be investigated analytically in $d=2$ dimensions \cite{David:1984tx, Kazakov:1985ea, Boulatov:1986jd} and by using numerical Monte Carlo methods when $d \geq 3$ \cite{Agishtein:1991ta,Boulatov:1991hg,Ambjorn:1991wq, Ambjorn:1991pq, Agishtein:1991cv, Catterall:1994pg}.
A natural question arises  if in the path integral \rf{Zcont} and consequently in the partition function \rf{Zdiscr} one should  only integrate/sum  over geometries of some chosen topology  or one should also include a summation  over various topologies, and if so, which topologies should be taken into account. The problem of summing over all possible topologies seems  to be ill-defined  even in two (spacetime) dimensions, where  
 the number of inequivalent geometries increases factorially with the Euler characteristic. As a consequence, the path integral \rf{Zdiscr} is not even Borel summable, and  a theory may have
(infinitely) many non-perturbative versions.  
The situation is even worse in  dimension four, where the classification of all possible topologies does not exist  {(see \cite{book} for a review of all these issues)}. Therefore  one is forced to restrict the class of admissible spacetime topologies, and usually only one simple topology is taken into account (in DT this was typically a four-sphere $S^4$).\footnote{Note that DT is formulated in the Euclidean regime and such a choice is compatible with Hartle-Hawking no-boundary proposal \cite{HartleHawkingWF}.} Numerical simulations of DT in spacetime dimension four showed that there exist two distinct phases, none of which 
resembles the four-dimensional Universe (see \cite{Ambjorn:1991pq, Agishtein:1991cv, Catterall:1994pg} and more recent \cite{Ambjorn:2013eha, Coumbe:2014nea}). The phases are separated by a first-order phase
transition  \cite{Bialas:1996wu, deBakker:1996zx}, which makes the possibility of approaching a continuum limit unlikely, however authors of Ref. \cite{Laiho:2016nlp} claim that such a possibility exists if bare coupling constants are fine-tuned in a specific way. 
Simulations using different topologies ($S^3\times~S^1$ and $T^4$) led to {the same results  \cite{Bilke:1996zd}, as one would expect
since phase transitions usually are associated with bulk properties where global 
topological constrains are of little importance. It has been conjectured that the first 
order transition is associated with large fluctuations of the lattice version of the conformal
mode  \cite{Dasgupta:2001ue}.}
 
{The causal dynamical triangulation approach (CDT) tries to change the 
above situation by starting out with a path integral defined for geometries with 
Lorentzian signature and by  assuming  that one only integrates over causal
geometries (implemented in the strong form of assuming that the geometry is 
globally hyperbolic), thus allowing the introduction of a global proper-time 
foliation. Since a topology change of a spatial slice often is associated with 
causality violation one forbids such changes in CDT
and as a result the spacetime topology is a product $\cM =  \Sigma \times I$, where $I$ denotes a (proper-time) interval and $\Sigma$ the spatial surface. All these concepts
are linked to the Lorentzian signature of the geometry. What is unique for CDT (compared to
DT) is that even at the discretized level, i.e.\ even in the 
regularized path integral,
each of these geometries has an analytic continuation to a geometry with Euclidean
signature and in the Euclidean regularized path integral  \rf{Zdiscr} we sum over this class
of geometries. The restriction is entirely motivated from the Lorentzian starting 
point. With such a genuine new set of Euclidean geometries there is a chance 
that the phase diagram of the lattice theory is changed compared to the DT
phase diagram. And this is indeed what happens.}

In CDT the time direction is distinguished, and the number of (discrete) time steps is kept fixed in numerical simulations. Due to the imposed time foliation, the four-dimensional triangulation is constructed from two types of elementary building blocks, the $(4,1)$ simplex with $4$ vertices on spatial hypersurface in time $t$ and $1$ vertex in $t\pm 1$, and the $(3,2)$ simplex with $3$ vertices in $t$ and $2$ vertices in $t\pm 1$. The simplices are glued together along their three-dimensional faces, and local curvature is defined by the deficit angle around two-dimensional triangles.
Additionally, trajectories are assumed to satisfy the simplicial manifold condition, i.e. every (sub-)simplex with a given set of vertex labels appears only once. It is also assumed that lattice spacing in time and spatial directions may be different, which defines the asymmetry parameter $\alpha$, such that:
\beql{alpha}
a_t^2 = \alpha\  a_s^2 \ .
\eeq  
The Regge-Einstein-Hilbert action takes the following form \cite{Ambjorn:2001cv}
\beql{SRegge}
S_{R}=-\left(\kappa_{0}+6\Delta\right)N_{0}+\kappa_{4}\left(N_{(4,1)}+N_{(3,2)}\right)+\Delta \ N_{(4,1)} \  ,
\eeq 
where $ N_{(4,1)}$,  $ N_{(3,2)}$ and $N_0$ denote the total number of $(4,1)$ simplices, $(3,2)$ simplices and vertices, respectively, while
$\kappa_{0}$, $\Delta$ and $\kappa_4$ are three  dimensionless coupling constants dependent on the bare Newton's constant, cosmological constant and the asymmetry  parameter $\alpha$.  

{In CDT the topology of  spatial hypersurfaces $\Sigma$ is a choice. The same is the case in
classical GR,  where the choice of topology will affect the possible classical solutions.
Similarly, we can expect the choice of topology of $\Sigma$ to affect the possible
semiclassical solutions of CDT around which the geometry will fluctuate in 
the quantum theory. In fact, since no background geometry is put into the path 
integral in CDT, changing spatial topology is a very interesting test of 
the picture that dynamically an average suitable background geometry is found
around which there are relatively small quantum fluctuations. Further, since 
CDT starts out as a lattice theory (before the lattice spacing is sent to zero), 
there is even an additional possibility to effectively defy the topology put 
in by hand, since on the lattice topology is only an approximate concept: while 
changing topology dynamically in the continuum will often result in infinite 
derivative terms, on the lattice such terms will be finite, thus allowing a change in
topology. This could happen if the entropy of the ``wrong'' configurations is so 
large that it overcomes the action barrier for a topology change. The CDT studies 
performed so far provide an example of this. Until now the computer simulations
have been performed with the spatial topology being $S^3$, while periodic 
boundary conditions were imposed in the time direction for purely technical reasons. 
Thus the total topology of spacetime imposed 
was $\cM =S^3\times S^1$. However, in the so-called phase C (to be described below)   the system effectively arranged itself into a geometry of 
 topology  $\cM =S^4$ (up to lattice artifacts), provided time was chosen sufficiently large. 
 This would not be possible for 
smooth classical solutions. However, once this has happened  and one takes the 
lattice spacing to zero, only the $S^4$ part will survive. The possibility of such 
dynamics makes it additionally  interesting to change the topology of $\Sigma$ and study
the corresponding CDT system.}

{The CDT studies with topology choice $\cM = S^3\times S^1$ have led to} 
the discovery of four distinct phases of geometry \cite{Ambjorn:2005qt,Ambjorn:2010hu,Ambjorn:2014mra,Ambjorn:2015qja}, including a physical phase C where a 4-dimensional  universe is observed \cite{Ambjorn:2005qt,Ambjorn:2004qm} with semiclassical features closely resembling  de Sitter  space \cite{Ambjorn:2007jv,Ambjorn:2008wc} and quantum fluctuations governed by the minisuperspace action \cite{Ambjorn:2008wc,Ambjorn:2011ph}. The physical phase C is   separated from phase A by a first order phase transition  and seemed to be separated from phase B by a second  order phase transition line \cite{Ambjorn:2011cg,Ambjorn:2012ij}. The more recent studies showed the existence of a new "bifurcation" phase (D) in between   phases C and B \cite{Ambjorn:2014mra,Ambjorn:2015qja},  and most likely the recently discovered C-D phase transition line is second order  \cite{Coumbe:2015oaa}, thus allowing  for the possibility of taking continuum limit from within the physically
interesting phase C \cite{Ambjorn:2014gsa,Ambjorn:2016cpa}.

In this work we investigate the impact of spatial topology change on  CDT results. We analyze a system where the spherical topology $S^3$ is replaced by a toroidal topology  $T^3\equiv S^1 \times S^1 \times S^1$. For technical reasons we  still keep  time-periodic boundary conditions, and therefore the resulting spacetime topology is $\cM = T^3 \times S^1$.\footnote{Note that the direction of time is still well defined and time is treated differently than space, i.e. we distinguish between  spatial Cauchy surfaces and keep the number of time steps fixed. The set up is thus very different from a f DT system with $\cM = T^4 = S^1 \times S^1 \times S^1 \times S^1$.}

\section{Simulation details}

The (Wick rotated) partition function \rf{Zdiscr} defines a statistical field theory which  can be studied by  numerical Monte Carlo methods. The idea is to define a Markov chain in the space of  all admissible triangulations, where a triangulation $\cT$ is generated with probability 
\beql{ProbT}
\hat P(\cT)= \frac{1}{\cZ}e^{-S_R[\cT]} .
\eeq 
This can be done by updating triangulations by a series of local Monte Carlo moves. In four-dimensional CDT one typically uses 7 types of moves which are {\it causal}, i.e. they preserve both the chosen spatial topology $\Sigma$ and the global spacetime topology $\cM=\Sigma \times S^1$, and  {\it ergodic}, i.e. any triangulation of topology $\cM$ can be obtained from any other triangulation of topology $\cM$ by a sequence of moves (more details can be found in \cite{Ambjorn:2001cv}). In addition we require that simplicial manifold conditions be satisfied. 

One usually starts the numerical simulation from an arbitrarily  chosen simple triangulation  $\cT_{start}$ , and by performing the moves one evolves the system in Monte Carlo  time. The moves are accepted or rejected according to the {\it detailed balance condition} which ensures that after a large number of attempted moves (the so-called {\it thermalization} period) the system tends toward a stationary state where the probability of generating a triangulation $P(\cT)\to\hat P(\cT)$. 
The 7 moves which were used in numerical simulations for topology $S^3\times S^1$  preserve the topology and causality, and can be applied to any topology $\Sigma\times S^1$. The missing part is generating an initial triangulation
$\cT_{start}$ for the new spatial topology $\Sigma=T^3$.

\begin{figure}[H]
    \centering
    \includegraphics[width=0.65\linewidth]{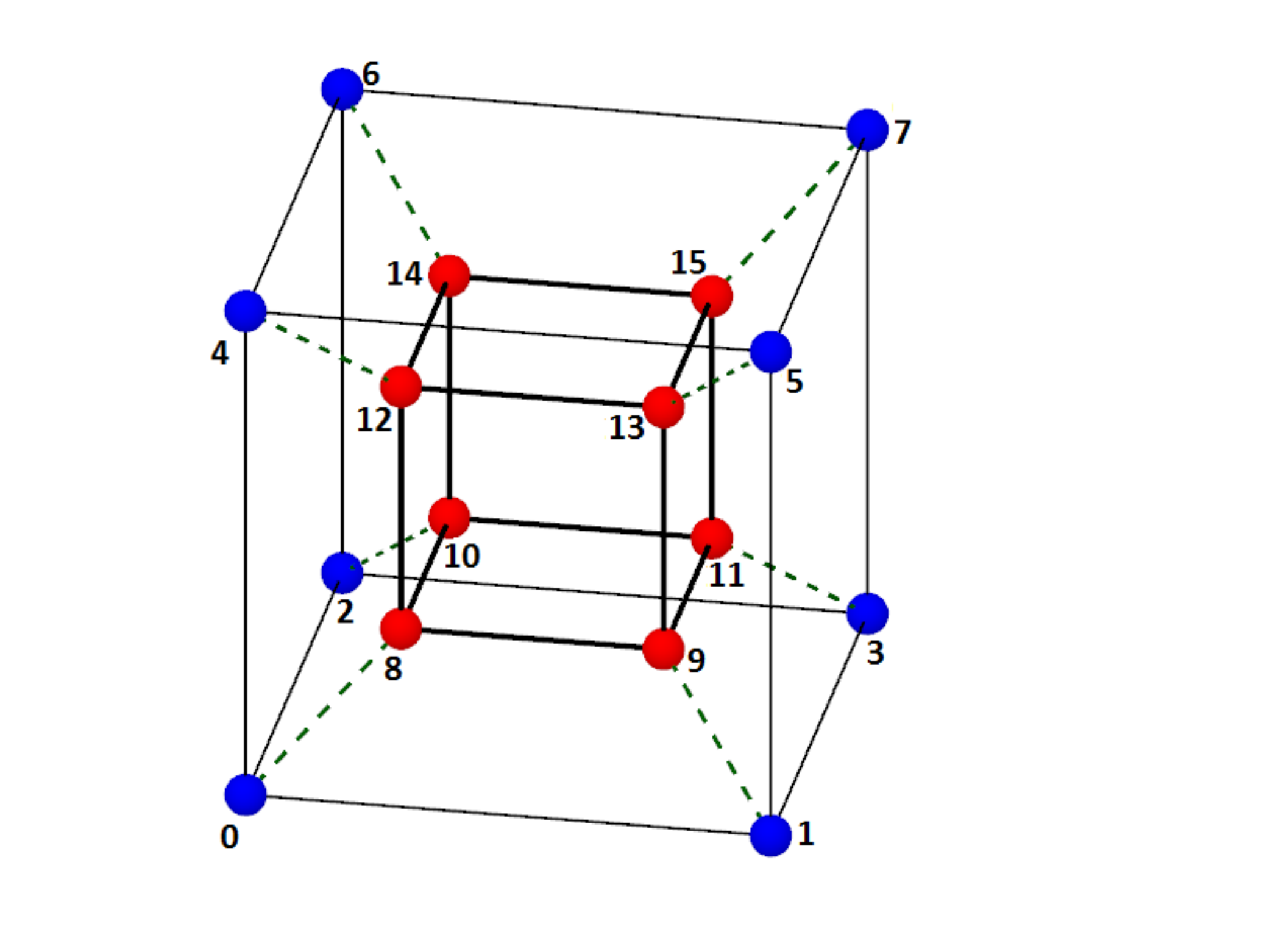}
    \caption{Visualization of a 4-cube.}
    \label{4-cube}
\end{figure}

We start by triangulating  a four-dimensional  hypercube into 16 simplices by using the  triangulation  method proposed by P.S. Mara~\cite{Mara}. The 4-cube has $16$ vertices, which can be seen in  the visualization in Fig. \ref{4-cube}, and we  assume that the blue vertices (labelled $0-7$) belong to the spatial Cauchy hypersurface (the 3-cube) in time $t$ and the red vertices (labelled $8-15$) belong to the 3-cube in time $t+1$. In the triangulation one obtains ten simplices with $4$ blue vertices and $1$ red vertex or vice-versa, all together being the $(4,1)$ simplices:  
\begin{center}
$\{0,1,2,4,8\},\{4,8,12,13,14\},\{2,8,10,11,14\},\{2,4,6,7,14\} ,\{1,2,4,7,14\},$\\
$\{1,8,11,13,14\},\{1,8,9,11,13\},\{1,4,5,7,13\},\{1,2,3,7,11\},\{7,11,13,14,15\} ,$\\
\end{center}
where the numbers in parentheses are vertex labels. One also obtains six simplices with $3$ vertices in one time slice and $2$ vertices in another (the $(3,2)$ simplices), parametrized by:
\begin{center}
$\{1,2,4,8,14\},\{1,4,8,13,14\},\{1,2,8,11,14\},$\\
$\{1,4,7,13,14\},\{1,2,7,11,14\},\{1,7,11,13,14\} .$ 
\end{center}

By gluing  the hypercubes together in all four directions we build a four-dimensional initial triangulation $\cT_{start}$, satisfying the regularity constraints. 
Only those d-dimensional (sub)simplices which share a d-1 dimensional face (d-1 simplex) can be glued together.
This can be done by joining together the hypercubes which are mirror images of each other in each direction.  As a result the number of  hypercubes in each direction  is even.  

\begin{figure}[H]
    \centering
    \includegraphics[width=0.8\linewidth]{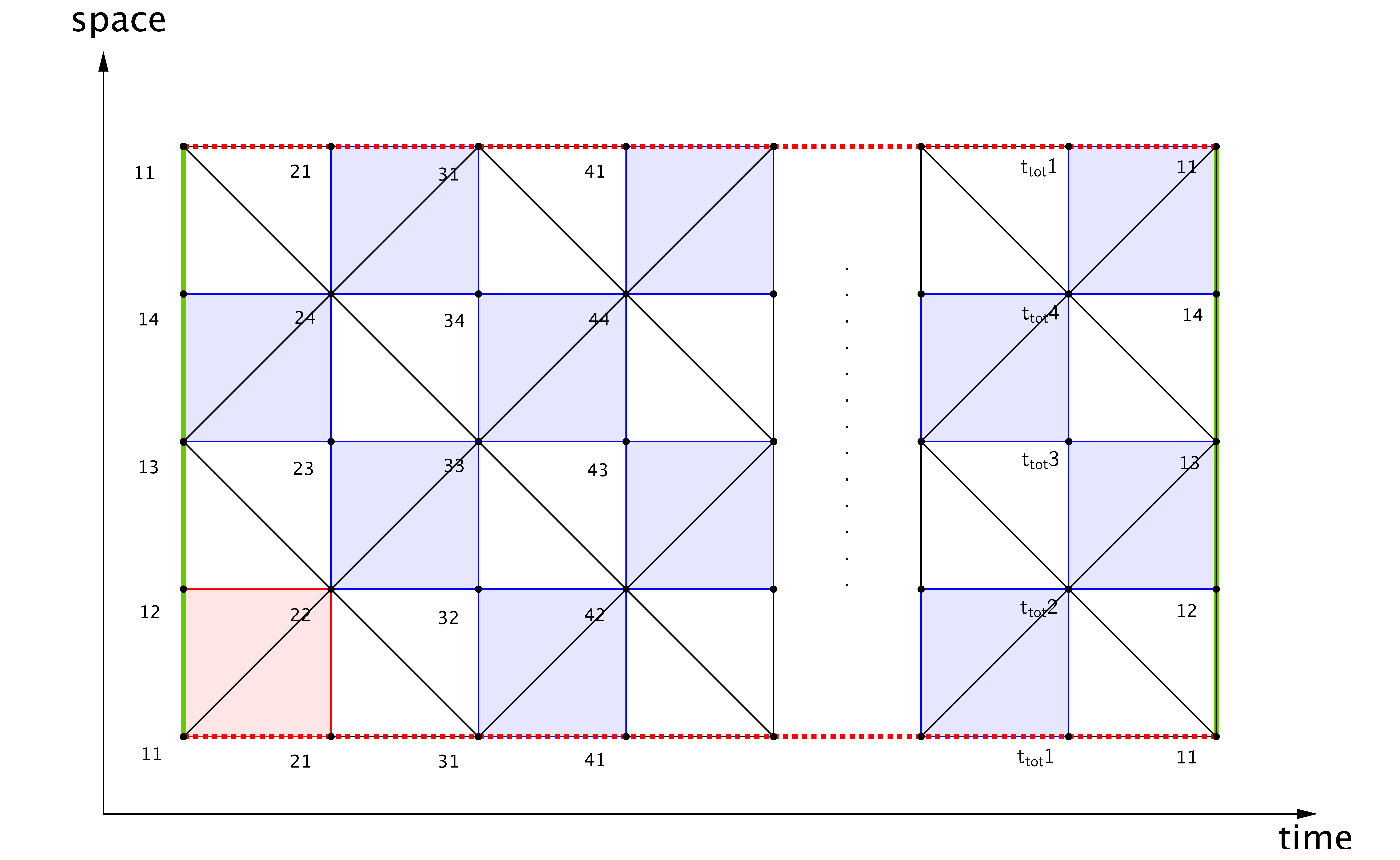}
    \caption{Two-dimensional visualization of the initial triangulation $\cT_{start}$. The starting cube (square) is colored in red. Colored and blank squares are mirror images of each other. The number of cubes (squares) in spatial direction is $4$, and the number in time direction is  $t_{tot}$ ($t_{tot} \geq 4$ and even).  The red and green edge lines are glued together, which is done by identifying the corresponding (sub)simplices  (vertices and links). Such boundary conditions are consistent with  toroidal topology of spatial slices resulting in global topology of $\cM = T^1 \times S^1$.}
    \label{2dimInitial}
\end{figure}

The configuration must be periodic in all $4$ directions, which can be done by identifying vertex labels of the "last" hypercube in each direction with that of the "initial" hypercube. Additionally we require that each d-simplex ($d=0,...,4$) with vertex labels $\{v_1,... ,v_{d+1}\} $  appear only once.   To fulfill these requirements one is forced to  use more than two hypercubes in each direction. We use $4$ hypercubes in each spatial direction and $t_{tot}$  hypercubes in the time direction ($t_{tot} \geq 4$ and even). A simplified two-dimensional visualization of the procedure is presented in Fig. \ref{2dimInitial}. The initial triangulation consists of $64 \cdot t_{tot}$ 4-cubes containing $1024 \cdot t_{tot}$ 4-simplices, out of which there are $384 \cdot t_{tot}$ $(3,2)$-simplices and $640\cdot t_{tot}$ $(4,1)$-simplices.  As a result each spatial slice is initially built from $320$ equilateral tetrahedra and has $64$ vertices.\footnote{By construction the number of $(4,1)$-simplices is twice the number of spatial tetrahedra. This is  because each spatial tetrahedron in time $t$ is a face of one $(4,1)$ simplex with the fifth vertex in $t+1$ and one such a simplex with the fifth vertex in $t-1$.}

It is worth  mentioning that the initial triangulation $\cT_{start}$  is not the {\it minimal} triangulation, i.e. the one containing the smallest possible number of (sub)simplices, but it is relatively easy to construct. By applying  Monte Carlo moves, we managed to shrink the triangulation to the one which  has only $90$ tetrahedra and  $15$ vertices in each spatial layer. It seems that the resulting spatial configuration is the smallest possible triangulation of a torus $T^3$ (by construction the moves do not allow for spatial topology change). We discuss this result in  detail in Appendix 1.  Note that the minimal toroidal triangulation is still much bigger than the minimal spherical configuration from our previous measurements, which consisted of just $5$ tetrahedra and had only $5$  vertices in each spatial layer. One may therefore expect that finite size effects  are substantial in current numerical studies. 

\section{The results}

In order to study  CDT quantum gravity numerically, one needs to define observables whose expectation values or correlation functions can be measured in Monte Carlo simulations. The idea is to probe the space of all possible triangulations with the probability given by \rf{ProbT}. As a result one obtains a sample of triangulations $\{ {\cal T}_1, {\cal T}_2, ..., {\cal T}_{N_{MC}} \}$ which can be used to estimate expectation values or correlation functions:
\begin{eqnarray}\label{correlsamp}
\langle{{\cal O}_1 ... {\cal O}_n} \rangle &= & \frac{1}{\cal Z} \sum_{{\cal T}}   {\cal O}_1({\cal T}) ...  {\cal O}_n({\cal T})    e^{- S_R[{\cal T}]}   \nonumber \\
&\approx& \frac{1}{N_{MC}}\sum_{i=1}^{N_{MC}}  {\cal O}_1({\cal T}_i) ...  {\cal O}_n({\cal T}_i)  \ . \end{eqnarray}
One then typically explores  the bare couplings parameter space  to check how the observables depend on the position in the parameter space. 
In principle, various phases can be identified and investigated using this method.
For four spacetime dimensions, CDT parameter space is spanned by three bare couplings: $\kappa_0, \Delta$ and $\kappa_4$. $\kappa_4$ plays the role of the bare cosmological constant, and the leading behaviour of the partition function \rf{Zdiscr} is
$$
Z \propto e^{(\kappa^c_4 - \kappa_4) N_4 } \quad , \quad N_4=N_{(4,1)}+N_{(3,2)},
$$
where $\kappa^c_4 = \kappa^c_4(\kappa_0,\Delta)$ is a critical value for which the theory becomes exponentially divergent. In numerical simulations we  
fix the total number of simplices, which in practice means that we approach a critical value $\kappa_4 - \kappa_4^c
\approx 1/N_4$
and thus we are left with a two-dimensional parameter space $(\kappa_0, \Delta )$. One can then check how the observables scale with increasing lattice volume $N_4$ and thus draw conclusions about the infinite volume limit. In CDT we typically control the lattice volume by introducing    an additional volume fixing potential to the bare action \rf{SRegge}. Here we use a quadratic volume fixing term
\beql{Svf}
S_{VF} = \eps \left( N_{(4,1)} - \bar V_{4}  \right)^2 
\eeq
with $\eps=0.00002$ controlling the amplitude of  volume oscillations around $\bar V_4$.

For the previously used spherical spatial topology the parameter space has been investigated in detail, which led to the discovery of four distinct phases \cite{Ambjorn:2005qt,Ambjorn:2010hu,Ambjorn:2014mra,Ambjorn:2015qja}. Here, with  toroidal spatial topology, we will focus on one particular point in the parameter space, namely $(\kappa_0=2.2, \Delta=0.6 )$, which in the spherical case was placed in the physical de Sitter phase C.\footnote{For the toroidal case we investigated several points $(\kappa_0, \Delta )$ of the bare parameter space located inside each of the previously discovered (spherical case) phases: A, B, C and D. Preliminary results show that  similar phases may exist in new topological conditions, however  it is not easy to distinguish them by using the methods  discussed here. One should also note that finite size effects are much stronger than for the spherical case, and thus one has to simulate with much higher lattice volumes to observe the differences. The results for other  points in the parameter space will be discussed elsewhere.}

\subsection{Spatial volume profile}

\begin{figure}[H]
    \centering
    \includegraphics[width=0.65\linewidth]{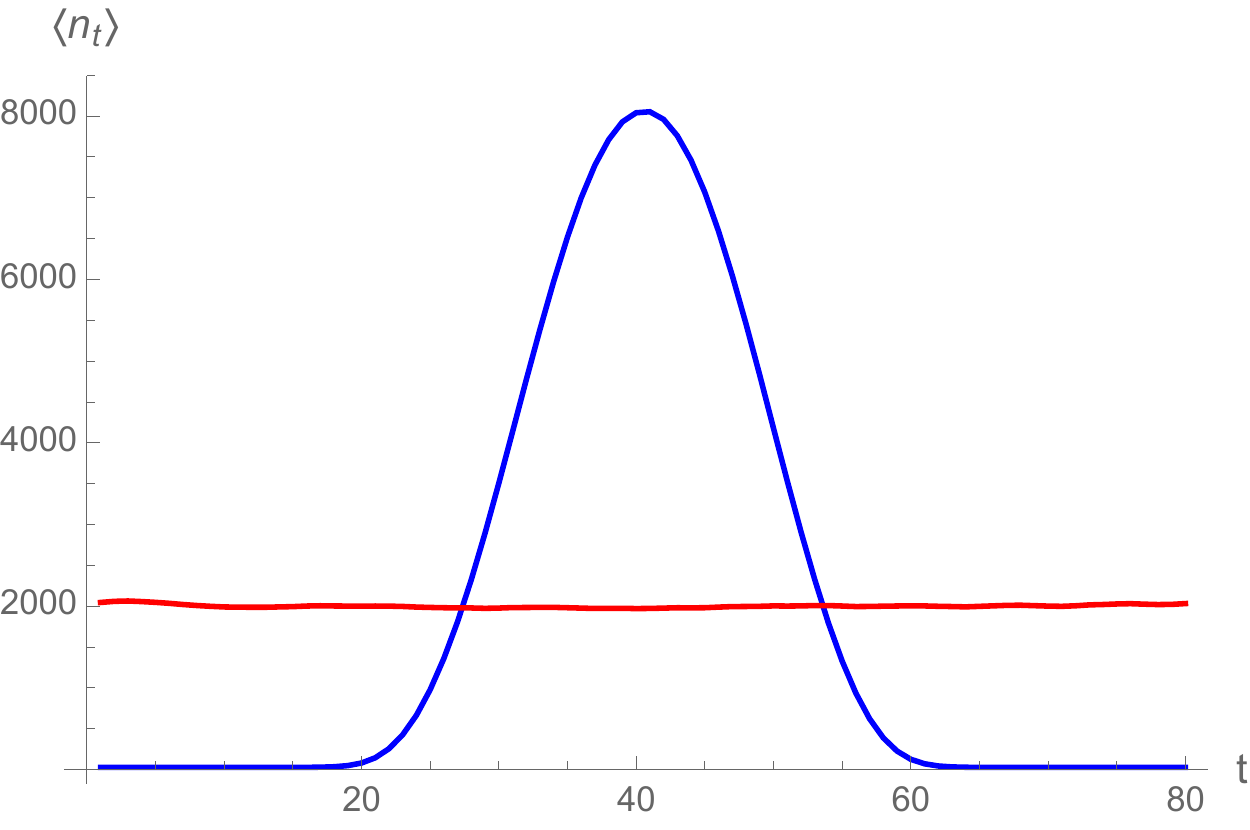}
    \caption{The spatial volume profile $\langle n_t \rangle$ measured for the toroidal (red line) and spherical (blue line) spatial topology, respectively. Data measured for $(\kappa_0=2.2 \, , \, \Delta=0.6)$.}
    \label{FigVolProf}
\end{figure}

An observable  investigated in this article is the {\it spatial volume}: 
\beql{Eqnt}
n_t \equiv N_{(4,1)}(t) ,
\eeq
where $N_{(4,1)}(t)$ is the number of $(4,1)$ simplices having  $4$ vertices in time $t$. This is by construction equal to twice the number of tetrahedra forming a spatial slice in  $t$ and, as all spatial tetrahedra are equilateral, the number is proportional to the physical 3-volume of a  Cauchy hypersurface in $t$. In numerical simulations one can measure  the average
\beql{avnt}
\bar n_t = \langle n_t \rangle ,
\eeq
called the {\it volume profile}, and the correlator
\beql{covmatrix}
\text{C}_{t,  t'} \equiv \langle(n_t-\bar n_t)(n_{t'}-\bar n_{t'}) \rangle ,
\eeq
called the {\it covariance matrix}.
For the spherical spatial topology, the  volume profile observed inside phase C had a blob structure which could be very well fitted with the $\cos^3$ function characteristic for the Euclidean de Sitter solution \cite{Ambjorn:2007jv,Ambjorn:2008wc} (see Fig. \ref{FigVolProf} - blue line). Now, for the toroidal topology, up to numerical noise, the spatial volume does not depend on $t$, and consequently the profile is  a constant line (see Fig. \ref{FigVolProf} - red line):
\beql{VolProf}
\bar n_t = \bar v \equiv \frac{\bar V_4 }{ t_{tot}} .
\eeq
{In  order to obtain a better understanding of the difference between $\bar{n}_t$
when $\Sigma = S^3$ and $\Sigma = T^3$ we will determine the 
effective action as a function of $n_t$. In the $S^3$ case this effective action was 
closely related to the classical minisuperspace action where only the scale 
factor of the universe was kept as a dynamical variable.}

\subsection{The effective action}

The partition function \rf{Zdiscr} of CDT can be rewritten in the following form:
\beql{Zdiscr2}
\cZ = \sum_{\cT}e^{-S_R[\cT]} = \sum_{\{n_t\}} \sum_{\cT_{\{n_t\} }} e^{-S_R[{\cT_{\{n_t\} }} ]},
\eeq
where the first sum on the right is over (all possible) spatial volume configurations $\{n_t\}\equiv (n_1, n_2, ..., n_{t_{tot}})$, and the second sum is over the subset of all  triangulations consistent with the spatial volume configuration  ${\cT_{\{n_t\} }}$, i.e. where $\forall t : \, N_{(4,1)}(t) = n_t$. 
{By performing the sum  $\sum_{\cT_{\{n_t\} }}$ on the rhs of \rf{Zdiscr2} we obtain
an effective action depending only on $\{n_t\}$: }
\beql{ZdiscrEff}
\cZ =  \sum_{\{n_t\}}  e^{-S_{eff}[n_t]},
\eeq
{This is  a minisuperspace theory of the scale factor 
($a(t) \propto n(t)^{1/3}$), but contrary to the ordinary minisuperspace theory 
it is exact, since we have integrated out the other degrees of freedom rather
than just dropping them. A priori it is not  clear that $S_{eff}[n_t]$ is useful. 
It could be very nonlocal. However that turned out not to be the case
when the spatial topology was $S^3$.  Measurements showed that $S_{eff}[n_t]$ 
could be described by a kinetic term and a potential which were closely  related 
to the terms  in the original Hartle-Hawking  minisuperspace model. Further, the measured minisuperspace action described (up to the numerical uncertainty inherited in 
the simulations) both the semiclassically observed background $\bar{n}_t$ and the 
quantum fluctuations of $n_t$ around this background. We will now employ the 
same methods as used for $S^3$ to determine the effective action when the 
space topology is $T^3$.}
{First we observe that the quantum fluctuations around $\bar{n}_t$ are relatively
small. Thus  it makes sense to expand $S_{eff}[n_t]$:} 
\beql{Ssemi}
S_{eff}[n_t] = S_{eff}[\bar n_t+\delta n_t]  = S_{eff}[\bar n_t] + \delta n_t \left. \frac{\partial^2 S_{eff}}{\partial n_t \partial n_{t'}} \right|_{\bar n_t}  \delta n_{t'} + o[\delta n_t ^3]
\eeq
and the effective propagator  is given by the (inverse) covariance matrix
\beql{Propagator}
\left. \frac{\partial^2 S_{eff} } { \partial n_t \partial n_{t'}} \right|_{\bar n_t} = C_{t , t'}^{-1} \ .
\eeq
Hence the covariance matrix measurement enables us to verify any conjectured 
form of the effective action. 
Based on the form of the effective action when topology of space is $S^3$ 
we make the following Ansatz: 
\beql{Seff}
  S_{eff}= \sum_t  \left( \frac{\Big( n_t-n_{t+1}\Big)^2}{\Gamma(n_t+n_{t+1}-2 n_0)} + V[n_t] \right) ,
\eeq
where $\Gamma$ and $n_0$ are constants. {We call the first term the kinetic term and the term $V[n_t]$ the potential.}  The constant volume profile \rf{VolProf} $\bar n_t = \bar v$ and the Ansatz \rf{Seff} imply that the inverse covariance matrix should have a simple tridiagonal form\footnote{Here we already subtracted the impact of the volume fixing potential \rf{Svf} which causes a  shift of all inverse covariance matrix elements by a constant $2\epsilon$. The volume fixing shift is also subtracted  from the measured (inverse) covariance matrix data.}
$$
C^{-1}=\scriptsize{
\left( {\begin{array}{*{20}{c}}
   2k+u & {-k} &0  & \cdots  &0 & -k \\
   { -k} & 2k+u & { -k} & 0 & \cdots   & 0 \\
   0 & { -k} & {2k+u} &  -k & \ddots    & \vdots \\
   \vdots & \ddots &  \ddots  &  \ddots  & \ddots   & -k \\
 -k&0 & 0  & \cdots  & -k& 2k+u \\
\end{array}} \right)} \nonumber \ ,
$$
with constant diagonal and sub/super diagonal elements defined by  the kinetic and potential coefficients $k$ and $u$, respectively 
\beql{Pcoeff}
k =\frac{1}{\Gamma(\bar v - n_0)} \quad, \quad u = V"[\bar v] \ .
\eeq

\begin{figure}[H]
    \centering
    \includegraphics[width=0.3\linewidth]{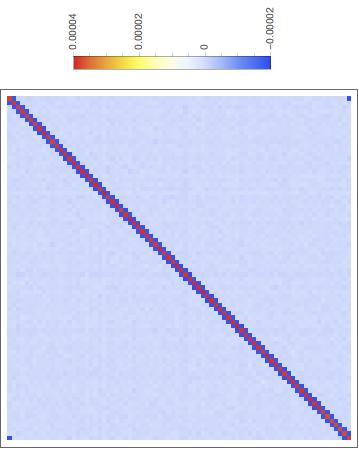}
  \includegraphics[width=0.6\linewidth]{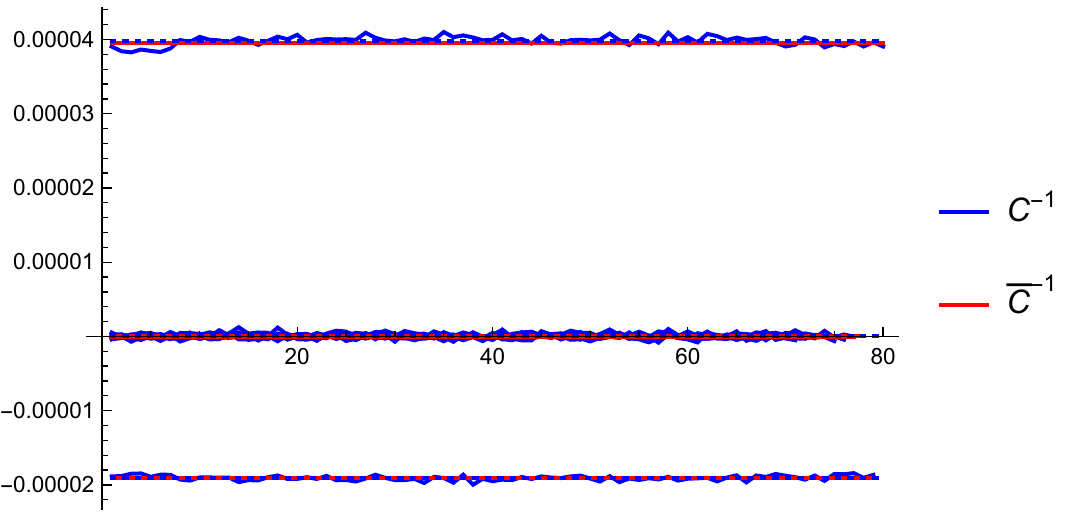}
    \caption{Left: The inverse covariance matrix $C^{-1}$ measured for $(\kappa_0=2.2 \, , \, \Delta=0.6)$, $\bar V_4=160000$ and $t_{tot}=80$. The structure is consistent with the assumed form of the effective action \rf{Seff}.  Right: The inverse covariance matrix diagonal (positive), sub/super-diagonal (negative)  and other matrix elements (close to zero). The $C^{-1}$ matrix elements are plotted as blue solid lines and the blue dashed lines are the averages (obtained by the method called "first invert then average" described in the text). The red lines are matrix elements obtained  by using the method "first average then invert" as described in the text. The results of the two methods are very similar and cannot be optically distinguished.}
    \label{FigInverseMatrix}
\end{figure}

\begin{figure}[H]
    \centering
    \includegraphics[width=0.7\linewidth]{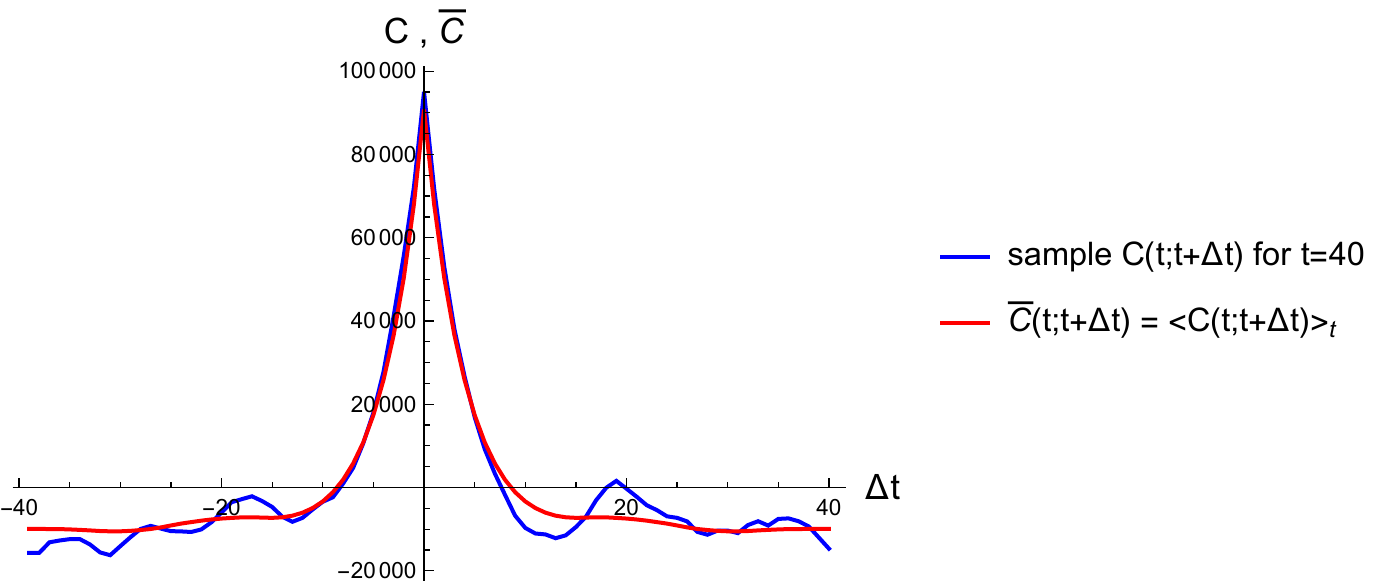}
    \includegraphics[width=0.4\linewidth]{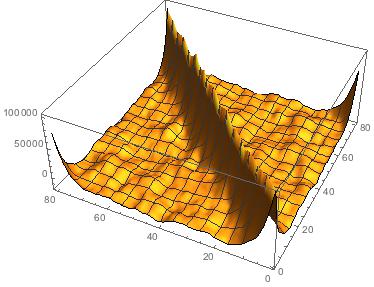}
    \includegraphics[width=0.4\linewidth]{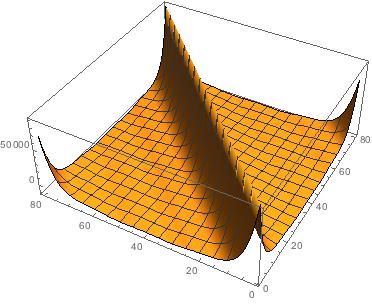}
    \caption{Illustration of the procedure "first average then invert" described in the text. The upper chart presents a sample row ($t=40$) of the measured covariance matrix $C_{t,t'}$  plotted as a function of $\Delta t=t-t'$ (blue line) and the "averaged" covariance matrix  $\bar C_{t,t+\Delta t} \equiv \langle C_{t,t+\Delta t}\rangle_t$ (red line). The lower chart presents the covariance matrix before (left) and after (right) applying the averaging procedure.}
    \label{FigTwoMethods}
\end{figure}

\begin{figure}[H]
    \centering
    \includegraphics[width=0.9\linewidth]{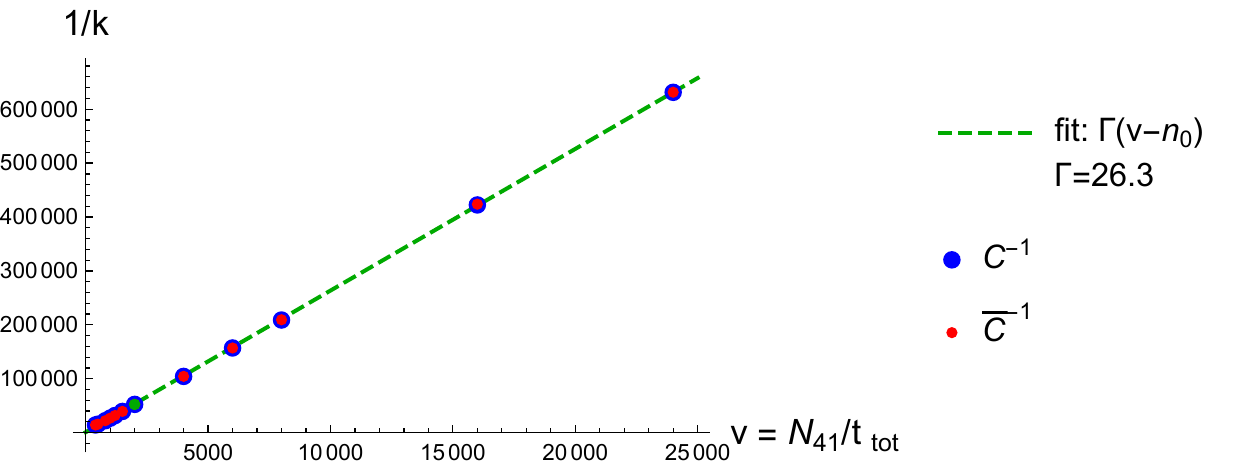}
    \caption{The (inverse) kinetic coefficients $k=-C^{-1}_{t,t+1}$ as a function of $\bar v = \bar V_4 / t_{tot}$ obtained for a collection of 12 measurements with $\bar V_4 = 80000,160000,240000$ and  $t_{tot}=10,40,160,200$. The linear relation is consistent with Eq. \rf{Pcoeff} and thus the Ansatz \rf{Seff}. Blue points were obtained by applying the procedure "first invert then average", and red points by the procedure "first average then invert", described in the text. The results of the two methods are (almost) identical. Green dots are the results obtained for $\bar V_4=160000$, $t_{tot}=80$  and thus $\bar v = 2000$. The dots are indistinguishable from the results for $\bar V_4=80000$, $t_{tot}=40$  resulting in the same $\bar v = 2000$.}
    \label{FigKinetic}
\end{figure}

\begin{figure}[H]
    \centering
    \includegraphics[width=0.9\linewidth]{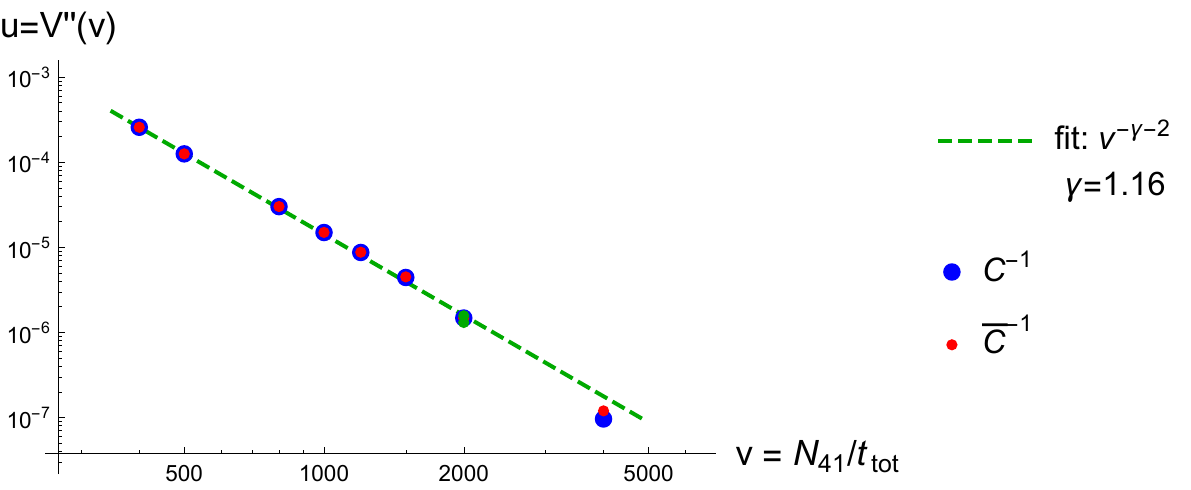}
    \caption{The potential coefficients $u=-C^{-1}_{t,t}-2k $ as a function of $\bar v = \bar V_4 / t_{tot}$ obtained for a collection of 12 measurements with $\bar V_4 = 80000,160000,240000$ and  $t_{tot}=10,40,160,200$ (data for $\bar v > 5000$ were skipped as they were indistinguishable from numeric noise). The plot is in the log-log scale and the dashed line is consistent with  a potential $V[\bar v] = \mu \, \bar v ^{\, -\gamma}$ - see Eq. \rf{Pcoeff} - and the best fit is for $\gamma=1.16\pm 0.02$. Blue points were obtained by applying the procedure "first invert then average", and red points by the procedure "first average then invert". The results of the two methods are (almost) identical. Green dots are the results obtained for $\bar V_4=160000$, $t_{tot}=80$  and thus $\bar v = 2000$. The dots are indistinguishable from the results for $\bar V_4=80000$, $t_{tot}=40$  resulting in the same $\bar v = 2000$.}
    \label{FigPotential}
\end{figure}

The covariance matrices measured for the point ($\kappa_0=2.2,\Delta=0.6$)  are indeed consistent with this  structure - see Fig. \ref{FigInverseMatrix},  where  we show the   data measured for $\bar V_4 = 160000$ and $t_{tot} = 80$ giving $\bar v = 2000$. In order to verify if  numerical errors are under control, we used two independent procedures for determining the coefficients $k$ and $u$. In the first method we simply invert the measured covariance matrix, and take the average of  diagonal or sub/super diagonal elements:
\beql{ku1}
k = - \langle C^{-1}_{t , t+1} \rangle_t \quad ,  \quad u = \langle C^{-1}_{t , t} \rangle_t -2 k \ .
\eeq
Accordingly the method can be called "first invert then average". In the second procedure we assume that due to a uniform spatial volume distribution ($n_t$ independent of $t$) the real dependence of the covariance matrix is on $\Delta t = t-t'$ and not on $t$. As a result $\forall t$ the matrix elements $C_{t,t+\Delta t}$  are identical up to numerical noise. Based on this assumption we calculate the "averaged" covariance matrix   $\bar C_{t,t+\Delta t} \equiv \langle C_{t,t+\Delta t}\rangle_t$ (see Fig. \ref{FigTwoMethods}) and then invert it. By construction we get constant (independent on $t$) diagonal and sub/super diagonal elements, resulting in
\beql{ku2}
k = - \bar C^{-1}_{t , t+1}=\text{const.}  \quad ,  \quad u = \bar C^{-1}_{t , t}  -2 k = \text{const.}  
\eeq
The method can be therefore called "first average then invert". We checked that the results of both methods are very consistent - see Fig. \ref{FigInverseMatrix}, where the blue dashed line and the red line corresponding to the two methods cannot be optically distinguished, and also Figs. \ref{FigKinetic} and \ref{FigPotential}, where blue and red dots are obtained by using the two methods respectively. 

In order to check the dependence on $\bar v$ and thus to verify whether the denominator of the kinetic term in Eq. \rf{Seff} is linear and find what is the shape of the potential part, one is forced to use a collection of  lattice volumes $\bar V_{4}$ and time periods $t_{tot}$. The results of such an analysis are shown in Figs. \ref{FigKinetic} and \ref{FigPotential}, where we present the measurements for all  combinations of $\bar V_4 = 80000,160000,240000$ and  $t_{tot}=10,40,160,200$ resulting in 12 different  values of $\bar v$. For  a consistency check, this  also includes $\bar v=2000$ which is identical as for the previously discussed case $(\bar V_4=160000 ; t_{tot} = 80)$. We checked that the  $k$ and $u$ coefficients measured for such $\bar v$ do not depend on the particular choice   of $(\bar V_4 \, ;\, t_{tot} )$  (green points in Figs. \ref{FigKinetic} and \ref{FigPotential}). 

The results obtained for various $\bar v$ show
that  the (inverse) kinetic coefficients $k^{-1}$ are indeed consistent with the expected linear behaviour $\Gamma (\bar v - n_0)$, and that the best fit of the potential coefficients of form $u[\bar v] = \mu (\gamma ^2 + \gamma) \bar v^{-\gamma-2}$ gives $ \mu > 0 $ and $\gamma = 1.16\pm 0.02 $, resulting in the following  form of the effective action consistent with  Ansatz \rf{Seff}:\footnote{Note that in the potential part we have included a linear term $+\lambda \, n_t$. The term is not recorded in the measured covariance matrix data which depend  on second derivatives of the effective action only. We have chosen the '+' sign (with $\lambda>0$) based on different measurement methods which will be described in a forthcoming article. }
\beql{SeffFinal}
  S_{eff}= \sum_t  \left( \frac{\Big( n_t-n_{t+1}\Big)^2}{\Gamma(n_t+n_{t+1}-2 n_0)} + \mu \, n_t ^{-\gamma} + \lambda \, n_t \right) \ .
\eeq



Major difference in the effective action between spherical and toroidal case
is visible in the potential term.
In the former case, the potential term is proportional to $v^{1/3}$ 
and is exactly the same as in the minisuperspace model.
Such term is a consequence of a constant, positive intrinsic curvature of spherical geometry
and can be verified by a direct calculation of the action.

The minisuperspace model reduces degrees of freedom to the scale factor $a(t) \propto v(t)^{1/3}$ 
and the minisuperspace action is obtained by inserting a maximally symmetric metric
\[ \dd s^2  = \dd t^2 + a^2(t) \dd \Omega^2, \]
into the (Euclidean) Einstein-Hilbert action.
For a unit three-sphere, the line element in spherical coordinates is given by
\[ \dd \Omega^2 =  \dd x_2^2 + \sin^2 x_2 \dd x_3^2 + \sin^2 x_2 \sin^2 x_3 \dd x_4^2. \]
Calculation of the Christoffel symbols and subsequently the Riemann tensor gives the scalar curvature
$R = \frac{6}{a^2}( - \dot{a}^2 - a \ddot{a} +1)$.
Integration by parts leads to the minisuperspace action
\[ S = \frac{1}{16 \pi G} \int \dd t \int \dd \Omega \sqrt{g} (R - 2 \Lambda) \propto  \frac{1}{G}  \int \dd t (a {\dot a}^2 - \frac{\Lambda}{3} a^3+a) , \]
or equivalently in terms of the spatial volume observable to
\[ S = \int \dd t \left(\frac{1}{\Gamma}\frac{\dot{v}^2}{v} - \lambda v  + \mu v^{1/3} \right) . \]

On the other hand, for a toroidal geometry 
the metric is Euclidean $\dd \Omega^2 =  \dd x_2^2 + \dd x_3^2 + \dd x_4^2$
and the intrinsic curvature vanishes. 
The scalar curvature $R = \frac{6}{a^2} (-\dot{a}^2 - a \ddot{a}) $
produces the toroidal minisuperspace action
\[ S \propto   \frac{1}{G}  \int \dd t (a {\dot a}^2 - \frac{\Lambda}{3} a^3) \quad \text{thus} \quad S = \int \dd t \left(\frac{1}{\Gamma}\frac{\dot{v}^2}{v} - \lambda v \right) . \]
Lack of the classical potential term might simplify observations of quantum corrections.


\section{Discussion}

We used the computer generated numerical data to measure the spatial volume profile and the covariance of spatial volume fluctuations in 3+1 dimensional CDT with toroidal spatial topology boundary conditions and to determine the effective action.
The form of the action observed for the 3+1 dimensional toroidal case \rf{SeffFinal}
\beql{ja1}
  S_{eff}^{(T^3)}= \sum_t  \left( \frac{\Big( n_t-n_{t+1}\Big)^2}{\Gamma(n_t+n_{t+1}-2 n_0)} + \lambda \, n_t  + \mu \, n_t ^{-\gamma} \right) 
\eeq
with $\gamma\approx 1.16$ can be compared with the minisuperspace action of the 3+1 dimensional  spherical case \cite{Ambjorn:2008wc,Ambjorn:2011ph}
\beql{ja2}
 S_{eff}^{(S^3)}= \sum_t  \left( \frac{\Big( n_t-n_{t+1}\Big)^2}{\Gamma(n_t+n_{t+1}-2 n_0)} - \lambda \, n_t + \mu \, n_t ^{1/3}  \right). 
\eeq
{The kinetic term present in both actions is a classical  term  in the sense that 
precisely such a term is present  in the minisuperspace reduction of the 
Einstein-Hilbert action both for the spatial topology $S^3$ and $T^3$. Interestingly, the numerical value of the effective parameter $\Gamma\approx 26.3$ measured in the point $(\kappa_0=2.2, \Delta=0.6)$ is, up to statistical precision, identical in both  cases.}
{The  potential term $ \mu\; n_t^{1/3}$ in \rf{ja2} is also a classical term as 
it is present  in the minisuperspace model when the spatial topology is $S^3$ and it is responsible 
for the semiclassical $S^4$-like background solution observed in the 
computer simulations. However, such a term
is not present in suitable minisuperspace reduction when the spatial topology 
is $T^3$ and we do not observe it in the computer simulations. 
The term we {\it do} observe, $n_t^{-\gamma}$, is numerically quite small and 
has the interpretation of a genuine quantum correction. The potential 
term is  purely due to quantum corrections.
It would be very  interesting to calculate analytically the exponent $\gamma$. }

\section*{Acknowledgements}

JGS and JJ wish to acknowledge the support of the grant DEC-2012/06/A/ST2/00389 from the National Science Centre Poland. 
JA and AG acknowledge support from the ERC Advanced Grant 291092 "Exploring the Quantum Universe" (EQU) 
and by FNU, the Free Danish Research Council, through the grant "Quantum Gravity and the Role of Black Holes". 
AG acknowledges support by the National Science Centre, Poland under grant no. 2015/17/D/ST2/03479. 
{In addition JA was supported in part by the Perimeter Institute of Theoretical Physics. Research at the Perimeter Institute is supported by the Government of Canada through Industry Canada and by the Province of Ontario through the Ministry of Economic Development and Innovation.}

\vspace*{24pt}


\renewcommand{\theequation}{A-\arabic{equation}}
\setcounter{equation}{0}

\begin{section}*{Appendix 1.  \ Minimal configuration of a three-torus}\label{App1}

In the presented setup, spatial slices are three-dimensional simplicial manifolds of a toroidal topology build of tetrahedra.
In this appendix we investigate the smallest triangulation of a three-torus,
i.e. possessing minimal number of vertices ($N_0$) and tetrahedra ($N_3$).
It is also a very interesting problem from a mathematical point of view.

By construction, CDT triangulations do not allow for distinct edges with the same endpoints.
Thus, the shortest loop consists of three vertices and three links (it is a triangle).
Naively, the minimal triangulation would be a Cartesian product of three such loops (one in each direction)  consisting of $3^3 = 27$ points.
Surprisingly, the smallest observed spatial slice has only $15$ points.
This is possible, because the loops in different directions interlace.
The found configuration consists of $N_0 = 15$ points,
$N_1 = \binom{15}{2} = 105$ links, $N_2 = 180$ triangles and $N_3 = 90$ tetrahedra.
It is a good candidate for the smallest triangulation of $T^3$ 
because it has a well defined structure and for $N_0 = 15$ only one combination of $N_1$ and $N_3$ was observed.

\subsection*{Layered structure}

The structure of the discovered minimal configuration is codified by links.
The number of links $N_1$ saturates the upper bound for $15$ points which means that each vertex is connected to each other.
There are only two types of links, (i) with coordination number (order) equal to $4$ and (ii) with coordination number equal to $6$.
Each point has identical vicinity shown in Fig. \ref{fig:min-cell}, it has $8$ outgoing links of order $6$ and
$6$ outgoing links of order $4$.
The latter links are marked with a thick red line and define the $x$, $y$ and $z$ axis (in both directions).
They also introduce a notion of layers, which can be identified with the $x$-$z$ planes.
Each of the three layers contains $5$ vertices and is enumerated by a discrete coordinate $y = 0, 1, 2$.

The layers are visualized in Fig. \ref{fig:min-layers},
the vertices have been relabelled so that the layers are
$L_0 = \{1, 2, 3, 4, 5\}$, $L_1 = \{6, 7, 8, 9, 10\}$ and $L_2 = \{11, 12, 13, 14, 15\}$.
Horizontal and vertical lines have coordination number $4$
and are drawn with a red line.
Taking a step to the right increases the $x$ coordinate by $1$ and the vertex label also by $1$ (modulo $5$),
going upward by one step increases the $z$ coordinate by $1$ 
but the vertex label by $2$ (modulo $5$) so that there are no two links with the same endpoints.
Each layer is periodic in $x$ and $z$ direction with period $5$.
The layers also form a structure with period $3$ ($L_0 \to L_1 \to L_2 \to L_0$).
One layer forms the smallest square grid with two non-equivalent and non-contractible loops.
There are also lines orthogonal to the plot which connect different layers (different $y$).
The layers visualized in Fig 2. should be viewed 
as placed on top of each other, so that e.g. vertex $1$ is connected to vertex $6$ by a link of order $4$.

\begin{figure}
	\centering
	\includegraphics[width=0.55\textwidth]{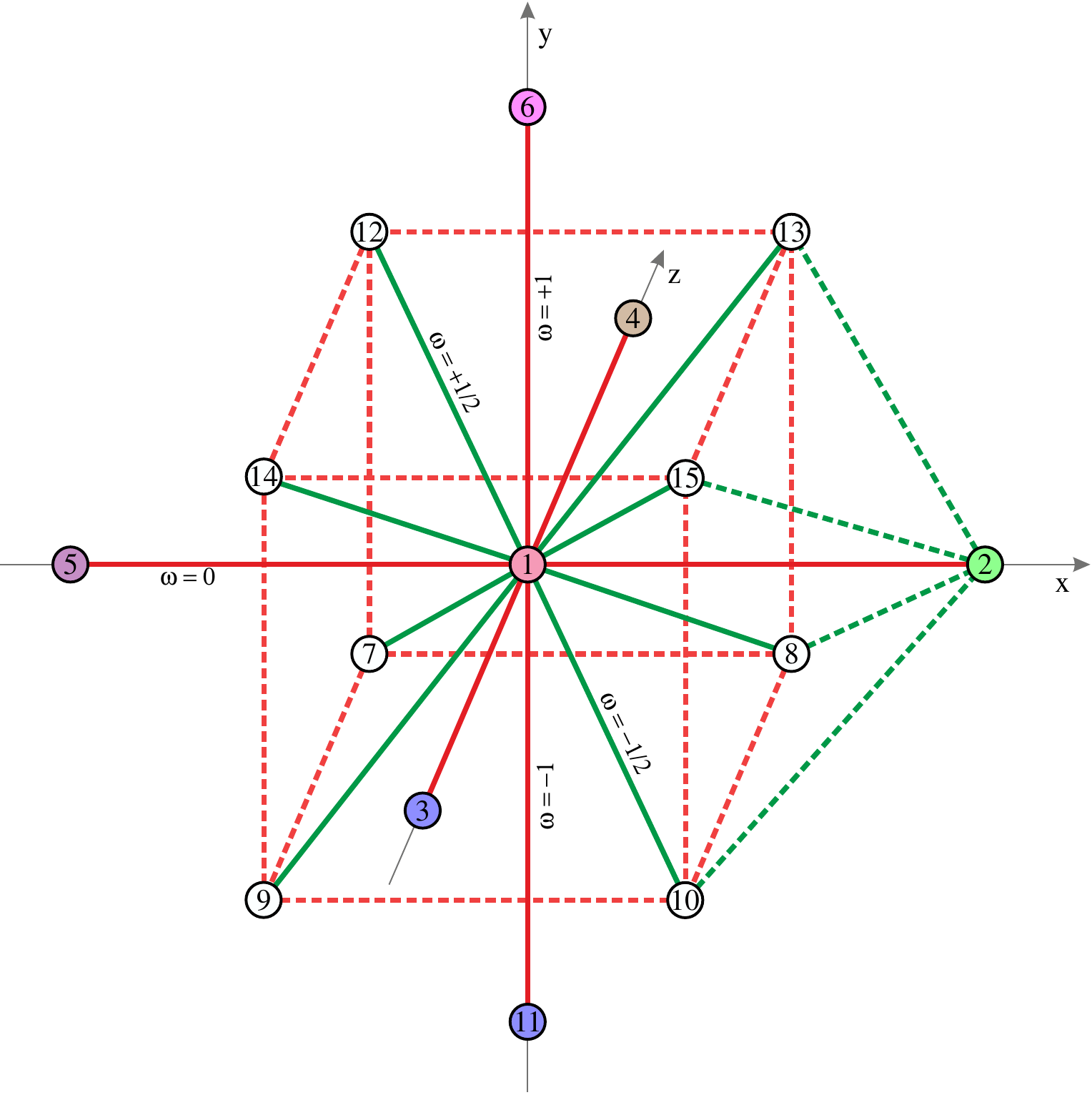}
	\caption{Neighborhood of vertex $1$. 
		The six outgoing links of order $4$ are marked with a thick red line, they determine the $x$, $y$ and $z$ directions.
		Vertices $2, 3, 4$ and $5$ lie in the same layer $L_0$ ($x$-$z$ plane) as vertex $1$.
		The eight outgoing links of order $6$ are marked with a thick green line.
		The structure for each point is identical. Note that each vertex is connected to each other.}
	\label{fig:min-cell}
\end{figure}

However, the whole picture is more complicated.
There are also links of order $6$, 
marked with a green line in Fig. \ref{fig:min-cell} and Fig. \ref{fig:min-layers},
which always connect different layers.
They change the $y$ coordinate by $\pm \frac{1}{2}$.
The layers are interlaced in such a way, that a layer $y = + 1$ (connected by links of order $4$) is at the same time at $y = -1/2$ (connected by links of order $6$).
For example, layer $L_2$ is simultaneously lying two steps above $L_0$, one step below or half step above.

\begin{figure}
\centering
\begin{tabular}{ccc}
\includegraphics[width=0.31\textwidth]{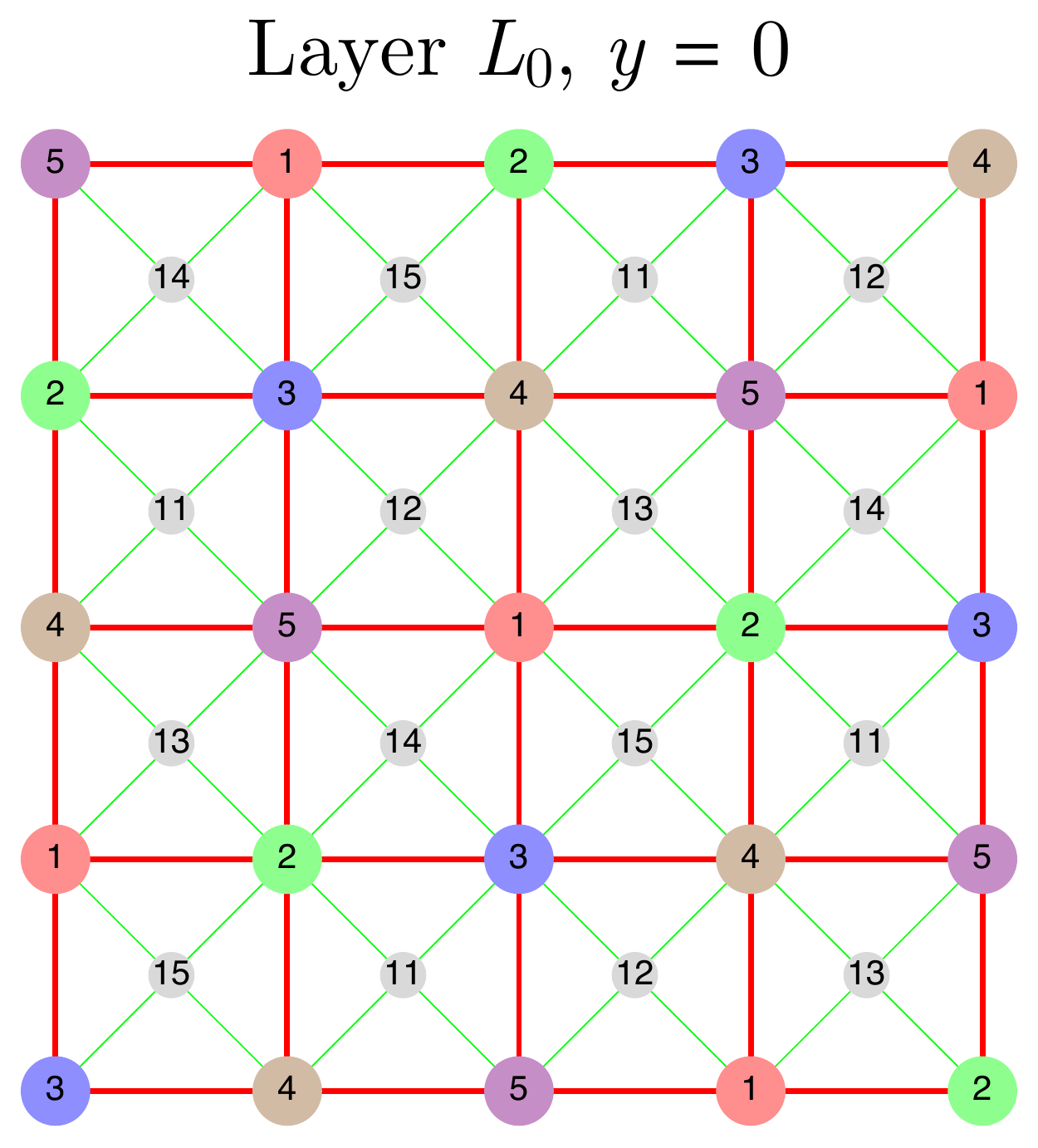} &
\includegraphics[width=0.31\textwidth]{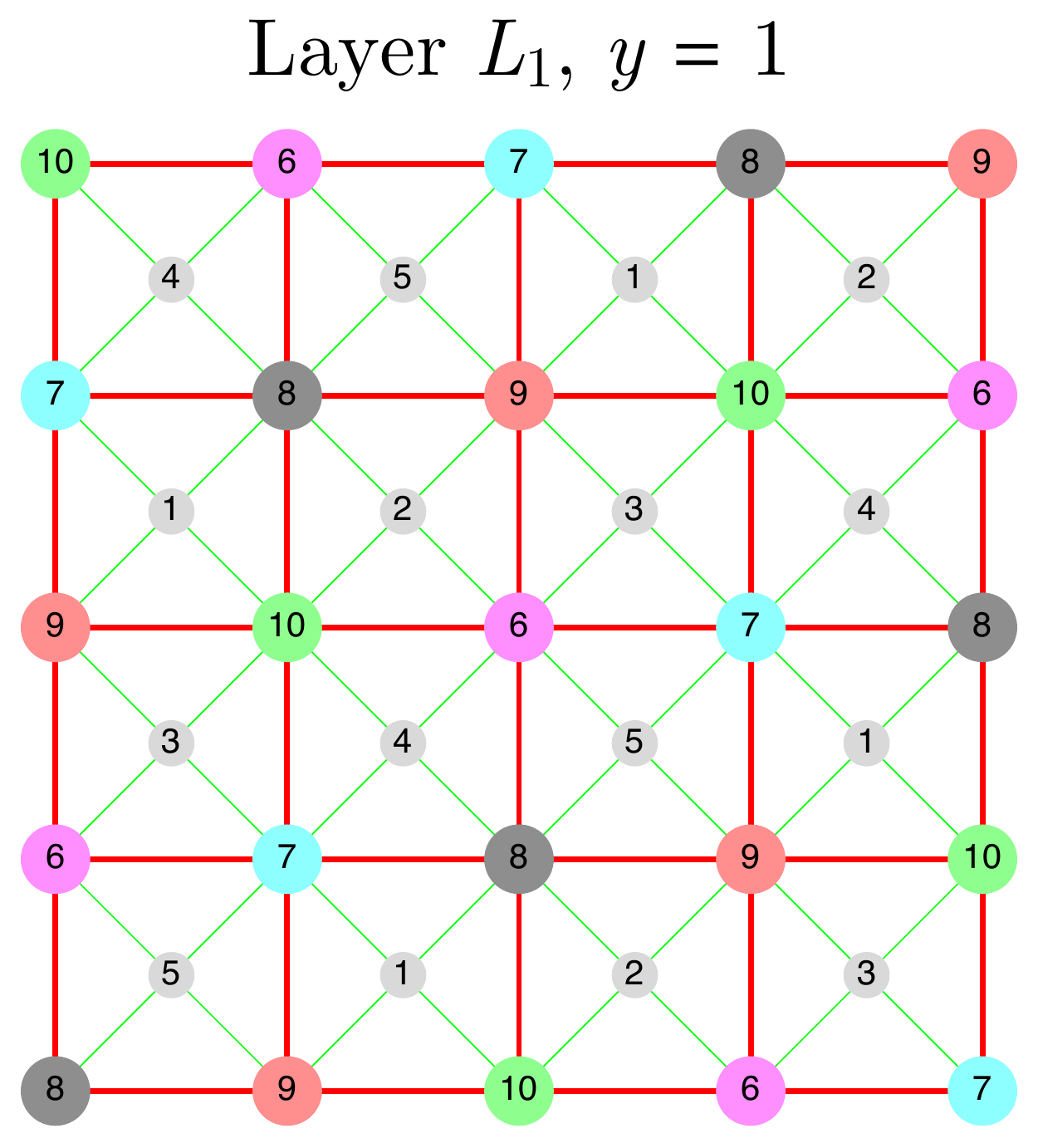} &
\includegraphics[width=0.31\textwidth]{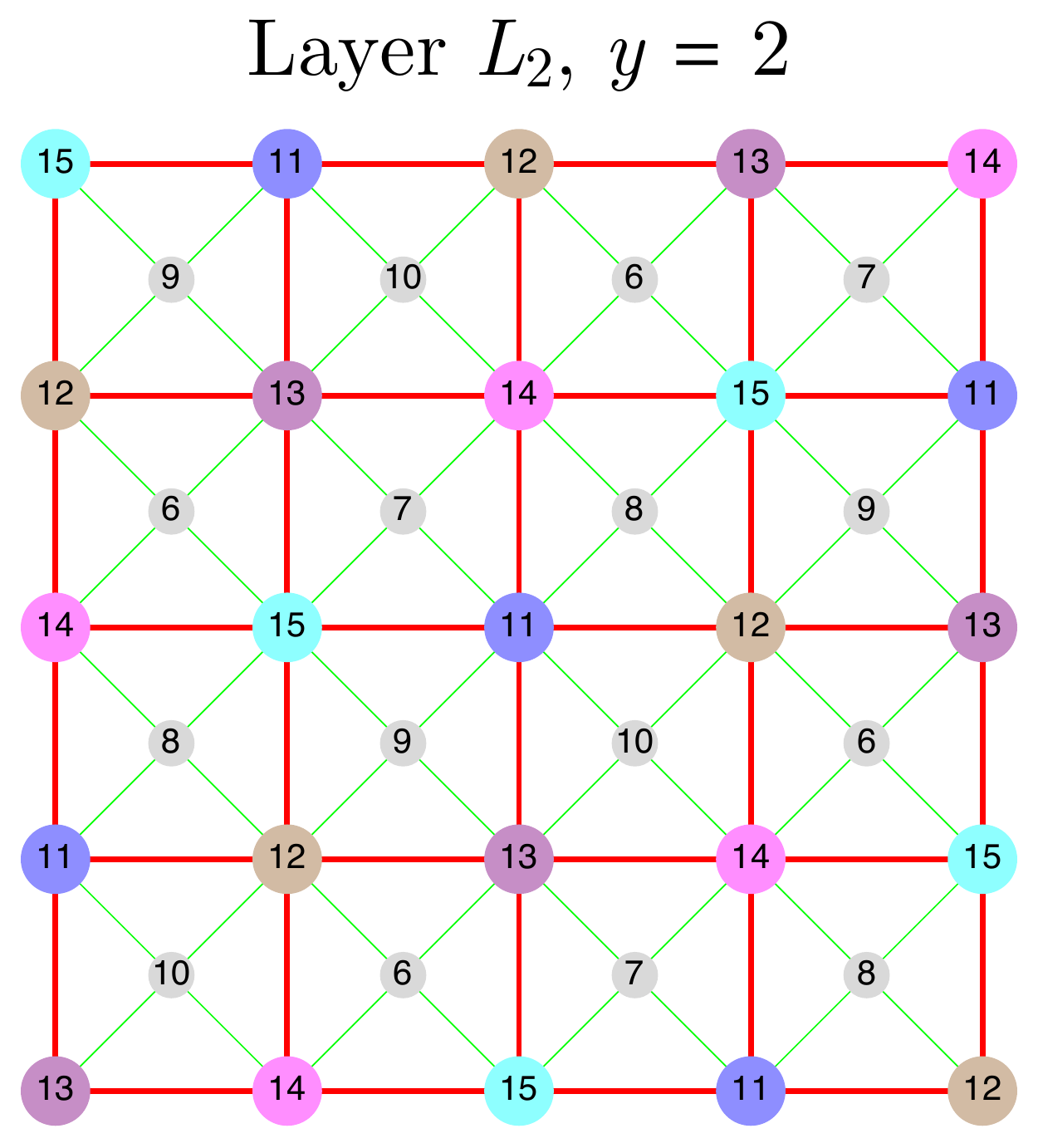}
\end{tabular}
\caption{Vertices of the minimal triangulation of $T^3$ can be divided into three disjoint layers of equal size.
The layers can be viewed as $x$-$z$ planes indexed by coordinate $y$.
Red lines denote links of order $4$ which connect points belonging to the same layer,
while green lines denote links of order $6$.
Links of order $4$ which connect adjacent layers are not visible.
Vertices lying in given layer are marked with color dots.
Gray dots denote vertices connected by links of order $6$ which belong to a layer shifted by half step in $y$ direction.
Each layer is periodic in $x$ and $z$ direction with period $5$.}
\label{fig:min-layers}
\end{figure}

\subsection*{Proof of toroidal topology}

To prove that the minimal triangulation indeed has a topology of a three-torus,
we have to show that there are three non-contractible and non-equivalent loops.
To demonstrate that all single winded loops can be split into three equivalence classes,
we assign a weight $\omega$ to each link.
Loops within one class can be continuously transformed into each other,
but not between different classes.
Link's weight basically corresponds to change in $y$ direction.
Because every loop can be continuously transformed 
into a path consisting only of edges present in the triangulation,
we restrict our considerations to such loops.

The weights are assigned according to following rules,
which are summarized in Table. \ref{tab:weights}.
Weight of a link depends only on its coordination number, orientation and layers it connects.
Edges that connect vertices lying in the same layer have weight $\omega = 0$
(horizontal and vertical red links in Fig. \ref{fig:min-layers}).
Links of order $4$ connecting layers in an increasing order (i.e $L_0 \to L_1 \to L_2 \to L_0$) have weight $\omega = +1$,
while links of order $6$ connecting layers in a decreasing order (e.g. $L_0 \to L_2$) have weight $\omega = +1/2$ 
(diagonal green links in Fig. \ref{fig:min-layers}).
Edges with opposite orientation have opposite weights.
Weight of a path is a sum of weights of links that build that path.

\begin{table}
	\centering
	\begin{tabular}{|c|c|c|}
		\hline\textbf{Order}&\textbf{Connection}&\textbf{Weight}\\ \hline \hline
		4	&  $L_y \to L_{y+1}$ & $ \omega = + 1 $ \\ \hline
		4 	&  $L_y \to L_{y-1}$ & $ \omega = - 1 $ \\ \hline
		4	&  $L_y \to L_{y}$ 	 & $ \omega =  0 $ \\ \hline
		6	&  $L_y \to L_{y-1}$ & $ \omega = + 1/2 $ \\ \hline
		6	&  $L_y \to L_{y+1}$ & $ \omega = - 1/2 $ \\ \hline
	\end{tabular}
	\caption{Weights $\omega$ of links with given order and endpoint layers. The orientation is important.}
	\label{tab:weights}
\end{table}

The clue of the proof is that any continuous transformation of a loop does not change its weight.
It can be inferred from Fig. \ref{fig:min-cell} that each tetrahedron has two links of order $4$, which don't meet, and four links of order $6$. 
An example of such tetrahedron is visualized in Fig. \ref{fig:min-trian} on the left.
This means that every triangle consists of two links of order $6$ with weight $\omega = \pm 1/2$ 
and one link of order $4$ with weight $\omega = -1, 0, +1$.
Thus every fundamental continuous transformation of a loop consists of changing one edge of a triangle into the two other or vice versa.
It is easy to show that in all cases the total weight is preserved.
\begin{description}
\item[Case I.] Let the red link connect layers $L_y \to L_{y+1}$. It has a weight $\omega = +1$.
Because the third point is connected by links of order $6$ it cannot lie in layer $L_y$ nor $L_{y+1}$
and has to belong to layer $L_{y+2}$.
Thus, the green lines connect layers $L_y \to L_{y+2}$ and $L_{y+2} \to L_{y +1}$ and both have weight $\omega = +1/2$ which gives in total $\omega = +1$.
\item[Case II.] Similarly, when the red link connects points in the same layer $L_y \to L_{y}$ it has a weight $\omega = 0$.
The third point has to be placed either in layer $L_{y-1}$ or $L_{y+1}$.
The green links are then $L_y \to L_{y+1}$ and $L_{y+1} \to L_{y}$ with total weight $\omega = -1/2 + 1/2 = 0$
or $L_y \to L_{y-1}$ and $L_{y-1} \to L_{y}$ with total weight $\omega = +1/2 - 1/2 = 0$.
\end{description}
Taking the orientation properly into account we can prove all remaining cases (e.g. transformation of a red and a green link into a green link).

\begin{figure}
\centering
\includegraphics[width=0.3\textwidth]{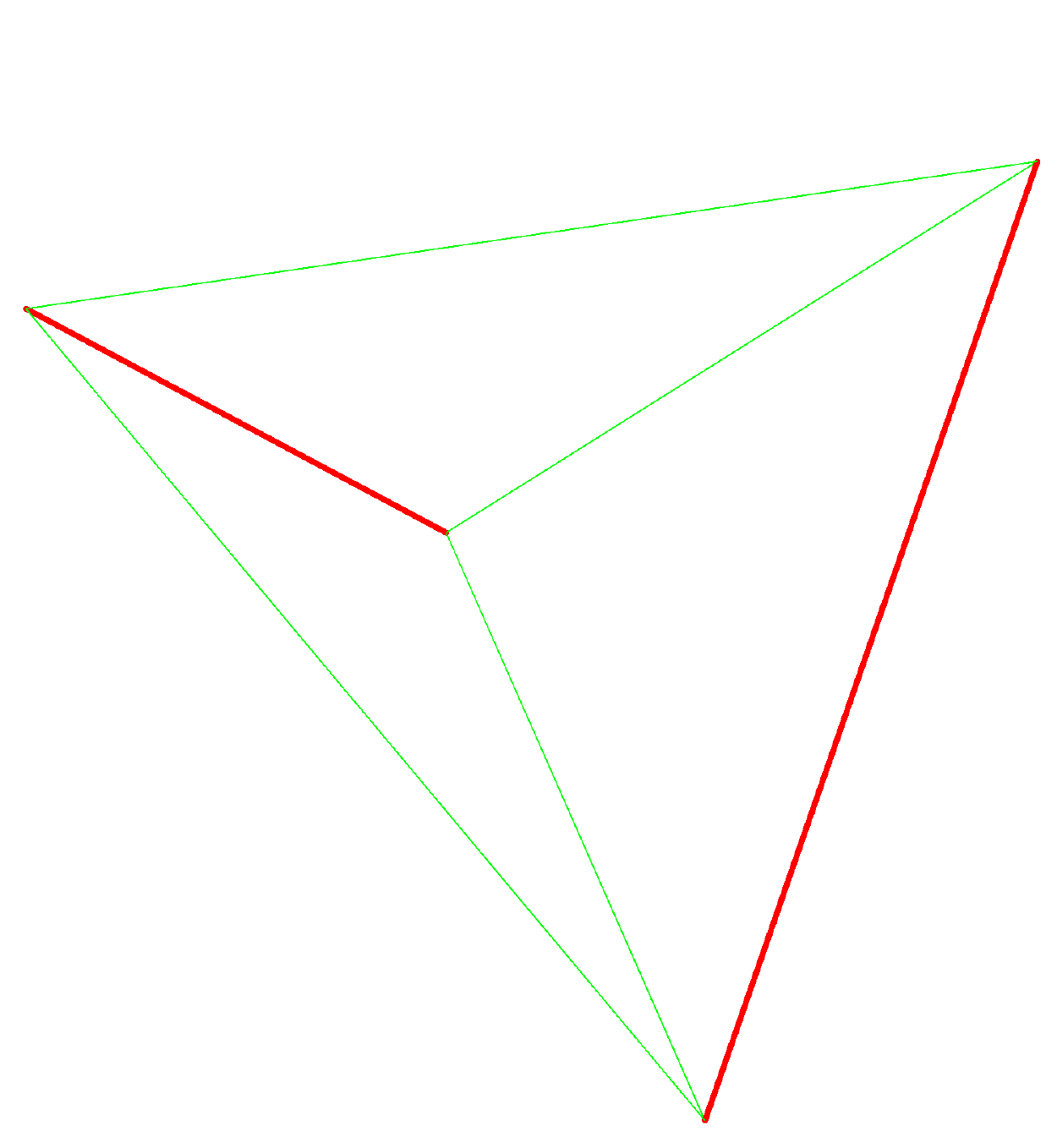}
\includegraphics[width=0.3\textwidth]{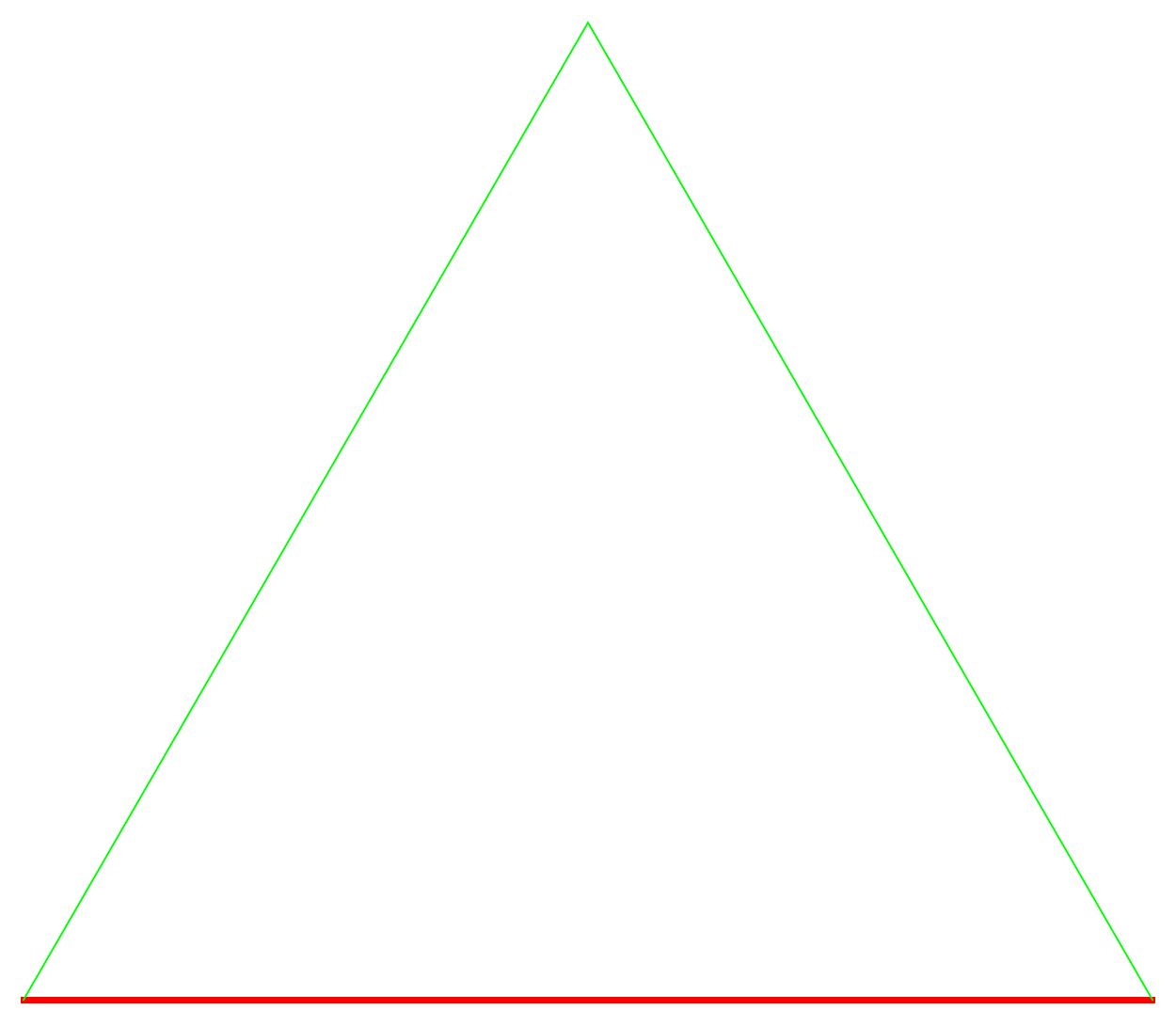}
\caption{
The figure shows a tetrahedron and a triangle (right) present in the minimal configuration.
Red line denotes a link of order $4$, green line denotes a link of order $6$.
All tetrahedra and triangles have this form.}
\label{fig:min-trian}
\end{figure}

The next step is to identify the equivalence classes of loops.
Let us consider following representatives:\\
\begin{description}
\item[First loop.]
\[ \mathbf{1}(L_0) \xrightarrow[\omega = 0]{4} \mathbf{2}(L_0) \xrightarrow[\omega = 0]{4} \mathbf{5}(L_0) \xrightarrow[\omega = 0]{4} \mathbf{1}(L_0)\]
Numbers in bold are labels of vertices belonging to the path, in parentheses are vertex layers.
Arrows denote links with endpoints on sides, the numbers above are link orders.
The weight can be derived from rules in Table \ref{tab:weights}.
This loop is completely embedded in layer $L_0$ and its total weight is $\mathbf{\omega = 0}$.
All contractible loops of length three form a triangle present in the triangulation.
Because there are no triangles composed exclusively of links of order $4$, this loop is non-contractible.
\item[Second loop.]
\[ \mathbf{1}(L_0) \xrightarrow[\omega = 1/2]{6} \mathbf{13}(L_2) \xrightarrow[\omega = 1/2]{6} \mathbf{10}(L_1) \xrightarrow[\omega = 1/2]{6} \mathbf{1}(L_0)\]
This loop passes through all layers in a descending order along links with coordination number $6$ (weight $+1/2$).
The total weight equals $\mathbf{\omega = 3/2}$. Because the weight is non-zero it cannot be a contractible loop.
\item[Third loop.]
\[ \mathbf{1}(L_0) \xrightarrow[\omega = 1]{4} \mathbf{6}(L_1) \xrightarrow[\omega = 1]{4} \mathbf{11}(L_2) \xrightarrow[\omega = 1]{4} \mathbf{1}(L_0)\]
This loop passes through consecutive layers along links of order $4$ (weight $+1$).
The total weight equals $\mathbf{\omega = 3}$ and because it is non-zero the loop is non-contractible.
It is also too short to be a loop of second type winded twice.
\end{description}
Moreover, the foregoing loops are non-contractible because they have length three but do not form a triangle present in the simplicial manifold.
Because they have different weights, they necessarily belong to separate equivalence classes.
This ends the proof of toroidal topology of the considered triangulation.
It is also noteworthy, that the minimal toroidal triangulation consists of $90$ tetrahedra,
which is much more than for a spherical topology ($5$ tetrahedra).

\end{section}

\bibliographystyle{unsrt}

\bibliography{Master}

\end{document}